\documentclass[twocolumn,showpacs,preprintnumbers,amsmath,amssymb]{revtex4}
\usepackage{graphicx}
\usepackage{dcolumn}
\usepackage{bm}

\begin{document}

\title{Decoherence of the two-level system coupled to a fermionic reservoir and the information transfer from it to the environment}
\author{Pei Wang}
\email{pei.wang@live.com}
\affiliation{Wangcunlvtukuang, 255311 Zibo, P. R. China}

\date{\today}

\begin{abstract}
We study the evolution of reduced density matrix of an impurity coupled to a Fermi sea after the coupling is switched on at time $t=0$. We find the non-diagonal elements of the reduced density matrix decay exponentially, and the decay constant is the impurity level width $\Gamma$. And we study the information transfer rate between the impurity and the Fermi sea, which also decays exponentially. And the decay constant is $k\Gamma$ with $k=2\sim 4$. Our results reveal the relation between information transfer rate and decoherence rate.
\end{abstract}


\maketitle

\section{introduction}

Coherence is the fundamental characteristics of quantum systems. The state of an isolated system is represented by a vector moving in Hilbert space. The state vector decides the mean value and probability distribution of observables. But in reality truly isolated systems do not exist. The systems we can observe are all open systems, which will exchange energy, matter and information with the rest of the world. The state of an open system is generally a mixed state and should be represented by a density matrix. Open systems show many different features from isolated ones. One of the most important is decoherence~\cite{zurek91,schlosshauer,leggett}, i.e., the nondiagonal elements of the density matrix in pointer state basis rapidly decay to zero. The decoherence can be modelled as a non-unitary process, which can be described by the master equation with a Lindblad decohering term~\cite{gorini76,lindblad76}. The decoherence marks the boundary between quantum world and classical world~\cite{zurek03}, and has been intensely studied because of its role in quantum measurement theory and quantum computation~\cite{zurek03,nielsen}. The two level systems coupled to spin bath~\cite{prokofev} and oscillator bath~\cite{leggett} are good examples for studying decoherence. Plenty of works have been contributed to this subject (for a review see Ref.~\cite{prokofev,zurek03,schlosshauer}). Recently the two level system coupled to fermionic bath~\cite{lutchyn,yamada} emerged before the explorers. In this paper we consider a simple fermionic bath: the resonant level model. 

Additionally there exists information transfer between an open system and its environment. The quantum measurement is realized due to the information transfer from the system to apparatus. Zurek noticed the relation between decoherence and information transfer from the system to the observer and environment~\cite{zurek03}. This relation is familiar to quantum mechanics community, as the well known state vector collapse in measurement. And the which-path experiments, which are designed to verify complementarity principle, give a straightforward picture between decoherence and information about particle positions. Generally speaking, any one unavoidably cause decoherence of a system when trying to get the information of it. 

The relation between information transfer and decoherence can be understood in the language of state vector. The system and environment compose an isolated system which is supposed to be in a pure state. If the composite system is all the time in a product state, the state of the system we concern is always pure and its evolution can be described by a unitary transformation. And the system and environment share zero information. Otherwise the composite system evolves into an entangled state. The evolution of system we concern is non-unitary, indicating decoherence process.
At the same time the system and environment exchange information, increasing their mutual information into a finite value. 

An open system is all the time transfering information outside and suffering decoherence. The dynamics of entropy entanglement (the half mutual information) has been studied in different one dimensional systems~\cite{eisler,chiara,eisert,calabrese}. While there is still little study about the relation between decoherence rate and the information transfer rate in a dynamical system. In this paper, we try to use an exactly solvable model to demonstrate this relation. 

\section{two level system coupled to fermionic reservoir}

The resonant level model is an exactly solvable model which has been widely used to describe the transport through quantum dot. The Hamiltonian of resonant level model is
\begin{eqnarray}\label{hamiltonian}
\hat H= \sum_k \epsilon_k \hat c^\dag_k \hat c_k + \epsilon_i \hat d^\dag \hat d + V\sum_k (\hat c^\dag_k \hat d +h.c.),
\end{eqnarray}
where $\hat c_k$ and $\hat d$ are the single electron annihilation operators in the lead and at the impurity site respectively. We set the Fermi energy of the lead to be zero. The resonant level model can be regarded as a two level system coupled to Fermi sea, if we use $|1\rangle$ to represent the state when the impurity site is occupied by an electron and $|0\rangle$ to represent the state when the impurity site is empty. We suppose at initial time the state of the impurity is $(\alpha |0\rangle + \beta |1\rangle) $. The decoherence will happen after we switch on the coupling between the impurity and the Fermi sea at time $t=0$. The pointer states are $|0\rangle$ and $|1\rangle $. The non-diagonal elements of the reduced density matrix of the impurity will decay to zero exponentially. 

The simple way to get the elements of the reduced density matrix at arbitrary time is to calculate the observables which can be related to them. We suppose the reduced density matrix is expressed as
\begin{eqnarray}\nonumber
\hat \rho_i(t) &=& \rho_{00}(t) |0\rangle \langle 0|+ \rho_{11}(t) |1\rangle \langle 1|\\ &&+ \rho_{01}(t) |0\rangle \langle 1 | + \rho_{10} (t) |1\rangle \langle 0 | .
\end{eqnarray}
At time $t=0$, we have $\rho_{00}(0)=|\alpha|^2$, $\rho_{11}(0)=|\beta|^2$ and $\rho_{01}(0)=\rho_{10}^*(0)=\alpha\beta^*$. We notice the annihilation operator at the impurity site can be expressed as $\hat d = |0\rangle \langle 1| \otimes \textbf{1}_F$, where $\textbf{1}_F$ is the identity operator on the subspace of the Fermi sea. Then it is easy to know
\begin{eqnarray} 
\langle \hat d(t) \rangle=\rho_{10}(t). 
\end{eqnarray}
Similarly we have $\hat d^\dag = |1\rangle \langle 0| \otimes \textbf{1}_F$ and $\hat d^\dag \hat d = |1\rangle \langle 1| \otimes \textbf{1}_F$, which indicate $\rho_{01}(t)=\langle \hat d^\dag(t) \rangle$ and $ \rho_{11}(t) = \langle \hat d^\dag (t) \hat d(t) \rangle$ respectively. 

We could calculate the expectation value of $\hat d(t)$ and $\hat d^\dag(t)\hat d(t)$ by non-equilibrium Green's function method or by diagonalizing the Hamiltonian~(\ref{hamiltonian}) (for details see Ref.~\cite{pei}). Here we follow the latter way, and first express the Hamiltonian in hybridization basis as
\begin{eqnarray}
 \hat H=\sum_s \epsilon_s \hat c^\dag_s \hat c_s ,
\end{eqnarray}
where $\hat c_s = \sum_k \frac{V}{\epsilon_s-\epsilon_k} B_s \hat c_k+B_s \hat d$. The transformation coefficient is $B_s=\frac{V}{\sqrt{(\epsilon_s-\epsilon_i)^2+\Gamma^2}}$, where $\Gamma=\rho\pi V^2$ is the linewidth of the impurity level and $\rho$ the density of states of the Fermi sea. The inverse transformation is $\hat d= \sum_s B_s \hat c_s$ and through this the time-dependent annihilation operator evaluates
\begin{eqnarray}
 \hat d(t)= e^{-i\epsilon_i t-\Gamma t} \hat d + V \sum_k \frac{e^{-i \epsilon_k t}- e^{-i\epsilon_i t-\Gamma t}}{\epsilon_k-\epsilon_i +i\Gamma} \hat c_k.
\end{eqnarray}
For simplicity we set $\hbar=1$ throughout the paper. Considering the initial condition we then have $\rho_{10}(t)  = \alpha^* \beta e^{-i\epsilon_i t-\Gamma t}$, $\rho_{01}(t) = \alpha \beta^* e^{i\epsilon_i t-\Gamma t}$, and
\begin{eqnarray}\nonumber
 \rho_{11}(t) &=& V^2 \sum_k \frac{|e^{i(\epsilon_k -\epsilon_i)t}- e^{-\Gamma t} |^2}{(\epsilon_k-\epsilon_i)^2 +\Gamma^2} \theta(-\epsilon_k) \\ && + e^{-2\Gamma t} |\beta|^2 .
\end{eqnarray}
The non-diagonal elements $\rho_{01}(t)$ and $\rho_{10}(t)$ decay exponentially, and the decay rate $\displaystyle \frac{d |\rho_{01}(t)|}{dt}=\frac{d |\rho_{10}(t)|}{dt} \propto \Gamma e^{-\Gamma t}$. This is the effect of decoherence. Here the Fermi sea plays the role of environment.

\section{Information transfer rate}

Next we study the information transfer between the impurity and the Fermi sea. In an open system, such as the impurity in resonant level model, the information will broadcast to the environment. This is a stumbling block in quantum computation, while a necessary process in quantum measurement where our purpose is to get the information of the system. The mutual information between the system and the environment $I$ measures how much information is shared between them. Its time derivative is a good measure of the information transfer rate, which is defined in this paper as 
\begin{eqnarray}
  T = \frac{1}{2} \frac{d I}{dt}.
\end{eqnarray}
Here we take the factor $\displaystyle \frac{1}{2}$ because the information transfer both from the impurity to the Fermi sea and vice versa. The composite system of the impurity and the Fermi sea is in a pure state. According to Schmidt decomposition~\cite{nielsen} we know the mutual information $I$ is two times of the entropy of the impurity, which can be got from the eigenvalues of the reduced density matrix. We find at arbitrary time the information transfer rate is
\begin{eqnarray}
 T=\frac{d \lambda_+}{dt} (\ln \lambda_- - \ln \lambda_+),
\end{eqnarray}
where $\lambda_\pm =\displaystyle \frac{1\pm \sqrt{1-4 (\rho_{00} \rho_{11} - \rho_{01}\rho_{10})}}{2}$.

\begin{figure}
\begin{center}
\includegraphics[width=0.45\textwidth]{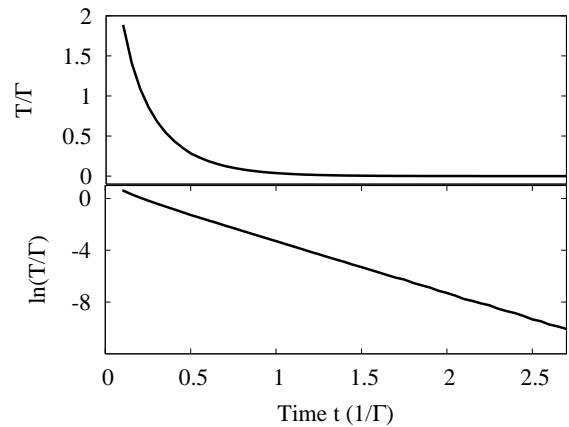}
\caption{Information transfer rate between the impurity and the Fermi sea. The impurity level is at the resonant position, i.e., $\epsilon_i=0$. The top figure shows the time-dependent information transfer rate. The bottom figure shows the logarithmic information transfer rate. We see clearly the exponential decay of $T$.}
\end{center}
\end{figure}
The non-diagonal elements of the reduced density matrix decay exponentially, so does the information transfer rate $T$ (see Fig.~1). The information transfer rate gradually decreases when the mutual information between the impurity and the Fermi sea increases to maximum. We employ $\Gamma$ as the unit of information transfer rate and $1/\Gamma$ the unit of time. The decay rate of the function $T(t)$ depends on $\Gamma$. At large time scale we approximately have $T\propto \Gamma e^{-k\Gamma t}$, where $k$ is a dimensionless constant. Recalling the fact that the decay rate of non-diagonal density matrix elements is proportional to $\Gamma e^{-\Gamma t}$, we get the relation between decoherence rate and information transfer rate: they both decay exponentially, and their decay constants are proportional to each other.

\begin{figure}
\begin{center}
\includegraphics[width=0.45\textwidth]{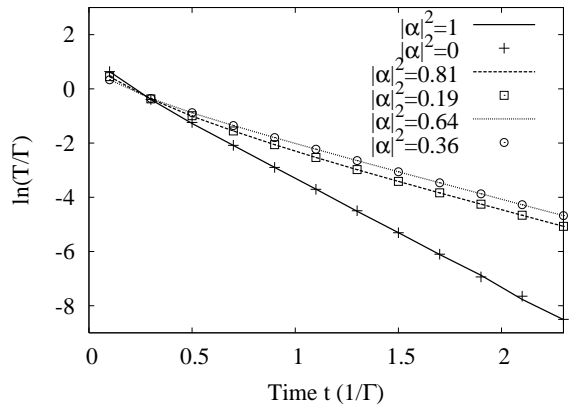}
\caption{Information transfer rate at different $|\alpha|^2$ when $\epsilon_i=0$. The decay constant of the information transfer rate depends on $|\alpha|^2$.}
\end{center}
\end{figure}
We find at large time scale the information transfer rate does not depend on the position of the impurity level $\epsilon_i$, but only depends on the initial condition $|\alpha|^2$. Fig.~2 shows the information transfer rate at different $|\alpha|^2$. The information transfer rate is proportional to $\Gamma e^{-k\Gamma t}$, where $k$ decreases from $4$ to $2$ when $|\alpha|^2$ decreases from $1$ to $0.5$, and increases back to $4$ when $|\alpha|^2$ decreases from $0.5$ to $0$.

\section{summary and conclusion}

We study the decoherence of an impurity coupled to a Fermi sea after the coupling is switched on at time $t=0$. The exponential decay of non-diagonal elements in the reduced density matrix of the impurity is observed. And the decay constant is found to be $\Gamma$. We also study the information transfer rate between the impurity and the Fermi sea. At large time scale the information transfer rate decays exponentially, and the decay constant is $k\Gamma$ with $k=2\sim 4$ depending on the initial state of the impurity (or $|\alpha|^2$). Our study reveals the relation between information transfer rate and decoherence rate.


\begin{references}

\bibitem{zurek91} W. H. Zurek, Physics Today {\bf 44}, 36 (1991).
\bibitem{schlosshauer} M. Schlosshauer, Rev. Mod. Phys. {\bf 76}, 1267 (2004).
\bibitem{leggett} A. J. Leggett, S. Chakravarty, A. T. Dorsey, M. P. A. Fisher, A. Garg, W. Zwerger, Rev. Mod. Phys. {\bf 59}, 1 (1987).
\bibitem{gorini76} V. Gorini, A. Kossakowski, and E. C. G. Sudarshan, J. Math. Phys. (N.Y.) {\bf 17}, 821 (1976)
\bibitem{lindblad76} G. Lindblad, Commun. Math. Phys. {\bf 48}, 119 (1976).
\bibitem{zurek03} W. H. Zurek, Rev. Mod. Phys. {\bf 75}, 715 (2003).
\bibitem{nielsen} M. A. Nielsen, and I. L. Chuang, Quantum Computation and Quantum Information (Cambridge Univ. Press, Cambridge, 2000).

\bibitem{eisler} V. Eisler, and I. Peschel, J. Stat. Mech., P06005 (2007).
\bibitem{chiara} G. De Chiara, S. Montangero, P. Calabrese, R. Fazio, J. Stat. Mech., P03001 (2006). 
\bibitem{eisert} J. Eisert, and T. J. Osborne, Phys. Rev. Lett. {\bf 97}, 150404 (2006).
\bibitem{calabrese} P. Calabrese, and J. Cardy, J. Stat. Mech., P04010 (2005). 


\bibitem{prokofev} N. Prokof'ev, and P. Stamp, Rep. Prog. Phys. {\bf 63}, 669 (2000).
\bibitem{lutchyn} R. M. Lutchyn, L. Cywinski, C. P. Nave, S. Das Sarma, Phys. Rev. B {\bf 78}, 024508 (2008).
\bibitem{yamada} N. Yamada, A. Sakuma, H. Tsuchiura, J. Appl. Phys. {\bf 101}, 09C110 (2007).
\bibitem{pei} P. Wang, M. Heyl, S. Kehrein, J. Phys.: Condens. Matter {\bf 22}, 275604 (2010).
\end{references}
\end{document}